\documentclass[fleqn,10pt]{wlscirep}

\usepackage{caption}
\usepackage{subcaption}

\title{Spatio-Temporal Patterns of the International Merger and Acquisition Network}

\author[1,*]{Marco Due\~nas}
\author[2]{Rossana Mastrandrea}
\author[3]{Matteo Barigozzi}
\author[4]{Giorgio Fagiolo}
\affil[1]{Department of Economics, International Trade and Social Policy, Universidad de Bogot\'a Jorge Tadeo Lozano, Bogot\'a}
\affil[2]{IMT School for Advanced Studies, Lucca}
\affil[3]{Department of Statistics, London School of Economics and Political Science, London}
\affil[4]{Istituto di Economia, Scuola Superiore Sant'Anna, Pisa }
\affil[*]{E-mail: marcoa.duenase@utadeo.edu.co (corresponding author)}

\begin{abstract}

This paper analyses the world web of mergers and acquisitions (M\&As) using a complex network approach. We use data of M\&As to build a temporal sequence of binary and weighted-directed networks for the period 1995-2010 and 224 countries (nodes) connected according to their M\&As flows (links). We study different geographical and temporal aspects of the international M\&A network (IMAN), building sequences of filtered sub-networks whose links belong to specific intervals of distance or time. Given that M\&As and trade are complementary ways of reaching foreign markets, we perform our analysis using statistics employed for the study of the international trade network (ITN), highlighting the similarities and differences between the ITN and the IMAN. In contrast to the ITN, the IMAN is a low density network characterized by a persistent giant component with many external nodes and low reciprocity. Clustering patterns are very heterogeneous and dynamic. High-income economies are the main acquirers and are characterized by high connectivity, implying that most countries are targets of a few acquirers. Like in the ITN, geographical distance strongly impacts the structure of the IMAN: link-weights and node degrees have a non-linear relation with distance, and an assortative pattern is present at short distances.
\end{abstract}

\begin{document}

\flushbottom
\maketitle

\thispagestyle{empty}

\section*{Introduction}

In the past two decades foreign direct investments (FDI) have become a major source of capital inflows for both developed and developing countries\cite{Qiu11}, with mergers and acquisitions (M\&As) as the dominant mode of FDI, representing more than 80\% of total national FDI\cite{WIR10}. 

In recent years, within the framework of a complex-network perspective, an increasing body of literature has been studying international trade\cite{sebo03, garlaschelli2005structure, serrano2006correlations, serrano2007patterns, fagiolo2008topological, fagiolo2009pre, fagiolo2010evolution, schiavo_etal_2010, de2011world, fagiolo2016oxford}, financial flows between countries, mostly considering transactions in equity securities, such as common stock and debt securities\cite{song2009statistical, joseph}, and, more recently, both financial and trade flows together\cite{schiavo_etal_2010, sgrignoli2014world, zhang2016}. The statistical analysis of these networks can complement traditional international trade and investment indicators, and can help explaining country growth and development patterns. 

Although the study of the determinants of M\&As has been of great interest to economists, only recently the attention has been focused on the topological properties of cross-border M\&As as a network of complex interactions (in terms of M\&As flows) between countries (nodes). Two recent contributions analysed cross-border investments for specific regions. S\'anchez D\'iez et al.\cite{sanchez2016fusiones} studied the role of Spanish investments in Latin America. They showed that Spain had played a central role in Latin American investments that recently decreased with the arrival of new investors. Using data from the OECD, Garas et al.\cite{garas2016} explored the properties and the link between the international migration network and the international FDI network (IFDIN), employing undirected network statistics and stocks of FDI. Using a gravity model, they found a strong relation among FDI and migration for those countries that are more central in the international migration network.

In this context, we introduce a pioneering study on the international M\&As network (IMAN) in space and time for a relatively long time period (1995-2010). We perform a statistical analysis for binary and weighted networks to provide evidence on the topological properties in different cross-sections. We study the architecture of the IMAN in different time-windows (from one to several months) in order to capture its prevailing or evolving patterns. Also, we use geographical distances between countries in order to study how geography affects the architecture of the network.

The main objective of our paper is to provide a descriptive characterization --as exhaustive as possible-- of the network properties of the IMAN in space and time, and a comparison with the international trade network (ITN), contributing to the understanding of international macroeconomics.

M\&As and trade can be understood as complementary ways of reaching foreign markets. Globalization can be characterized in terms of two unbundling dynamics: i) the first one derives from the fall in transportation costs and the removal of trade barriers, which spatially unbundled production from consumption, enabling international specialization and trade; and ii) the second one refers to the geographical separation of production, which starts taking place in different locations, as new technologies enable firms to relocate certain stages of the production process to other countries, increasing FDI and, in particular, M\&As\cite{baldwin}. In this way, the process of international trade and the process of relocating production have both changed during our period of study (and before). These phenomena create a natural framework to consider together the properties, structure, and behaviour of the IMAN and the ITN from a comparative perspective. We believe that our analysis allows to gain complementary insights on the relationships between countries through trade and M\&As. 

The space and time analysis of the IMAN are each one guided by two main motivations. Firstly, cross-border M\&As have experienced periods of impressive growth in the number of involved countries, the number of links, the number of acquisitions, and volumes. Their over-time dynamics has been characterized by a wave-like behaviour, specifically, with two waves in the period analyzed here. With the time analysis we are interested in studying whether the wave-like behaviour of the aggregate time series affects the architecture of the network. Our analysis allows us to conclude that, during the waves of M\&As, the IMAN is governed by an intensification/reduction of flows (intensive margin) rather than by the creation/destruction of links (extensive margin). This explains that several network statistics that characterize the IMAN are relatively stable over time.

Secondly, the effects of geography have been often neglected in international network analysis (for a review of different networks in which space has relevant implications see Barth{\'e}lemy\cite{barthelemy_spatial}). But distance certainly plays an important role at shaping the topological properties of the interactions among countries. Several authors have stressed that information asymmetries increase with distance creating a barrier to cross-border movements of capitals \cite{de_menil_1999,digiovanni_2005,portes_rey_2005}. As trade costs increase with distance, the simplest premise is that the decision to set up affiliates in foreign countries is positively affected by distance. In contrast, several empirical analysis show that the effect is negative. This evidence suggests that, in addition to trade costs, there might exist other sort of costs related to geographical distance. Investment interactions in the form of cross-border M\&As imply active management or control of the issuing companies, in contrast to passive financial investments. Thus, geographical distance might have a different effect on this kind of cross-border investments because higher transaction costs could be expected. We argue that these mixed findings might also be related with the existence of a non-linear effect of distance on M\&As. Then, the main objective of the spatial analysis is to study whether there are non-linear effects of geographical distance shaping the topology of the IMAN architecture. To do that, we analyse the network statistics restricting the interactions between country pairs to specific ranges of distances.

More in detail, our results are the following. The IMAN is very sparse, very concentrated in a few countries, and strongly target oriented, unlike the ITN. We find that the IMAN is a low density network characterized by a persistent giant strongly connected component with a few number of reciprocated links and with many weakly connected external nodes. The giant component is mainly composed by developed economies, which have more reciprocal investment relationships, high connectivity, and clustering.

We observe a high heterogeneity in the clustering of the IMAN. There are neither well established nor persistent hubs in the network. Interestingly, binary clustering coefficients are higher for countries known to have favourable legal and fiscal frameworks to attract FDIs and, particularly, for tax havens. On the other hand, mainly high-income and emerging market economies show high weighted clustering coefficients. This contrasts with the ITN, in which most clustered countries are quite stable over time\cite{fagiolo2009pre}. 

Despite the erratic dynamic evolution of M\&As, several network features do not change significantly over time and can be considered as stylized facts: i) the density is very low and reacts very little to changes in the number of links, ii) the size of the giant strongly connected component is around a half of the whole network, iii) correlations between node degree and node strength are very high and significant, and iv) clustering patterns are persistently unstable.

Finally, we show that there are strong non-linearities related to the geographical distance, with link-weight and node degree very high in the limit cases of short and relatively long geographical distances. An assortative pattern pattern emerges at short distances, i.e. at a regional level countries tend to be connected to other countries with similar node degrees and strengths. Interestingly enough, these non-linear patterns with distance are also present in the ITN\cite{abbate}.

\section*{Results}

\subsection*{Network topology}

\begin{figure}[!ht]
\centering%
\includegraphics[width=\textwidth]{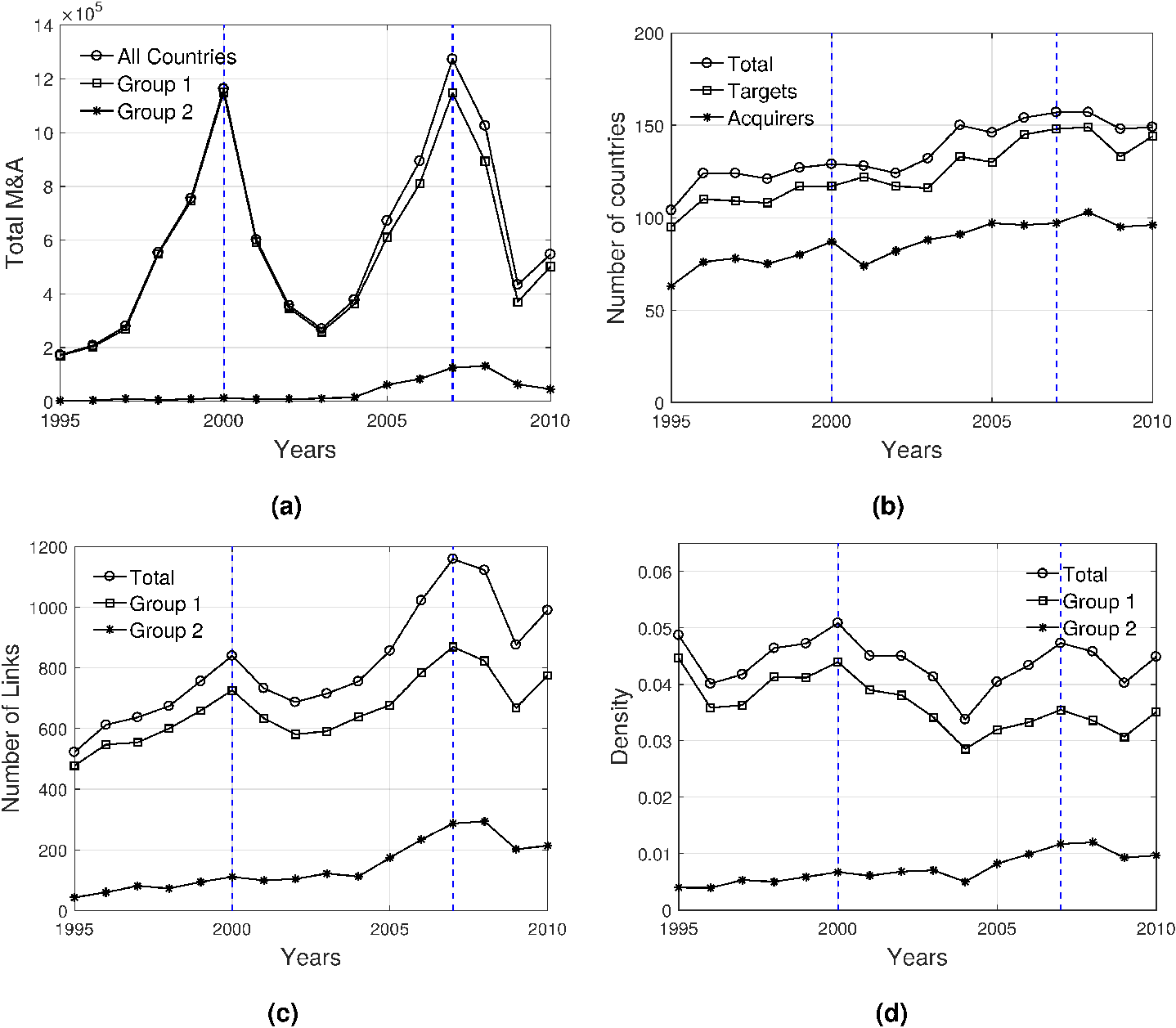}
\caption{\textbf{Network statistics.} \textbf{(a)} Evolution of total outflows of M\&As for all countries and for countries in different groups. \textbf{(b)} Evolution of the number of countries, targets, and acquirers. \textbf{(c)} Evolution of the number of links, and total number of acquisitions made by all countries and by different country groups. \textbf{(d)} Evolution of the network density and decomposition of total density by country groups. \textbf{Note:} Vertical lines indicate the peaks of the two waves. Group 1 includes countries that belong to the giant strongly connected component in all the years of the sample. Group 2 includes the set-difference between the union of all countries in the giant strongly connected component in any year and the members of group 1.} \label{fig:total_ma} 
\end{figure}

Capital investment worldwide has had remarkable changes since the mid 1990s. Figure~\ref{fig:total_ma}a shows the total amount of M\&As abroad cumulated over all countries. The time series of M\&As clearly depicts a so-called `wave-like' behaviour of cross-border M\&As, characterized by substantial variation over time, with some periods of rapid growth and other periods of rapid decline\cite{brakman}. In the period under study, we observe two waves: the first one between 1995 and 2003, and the second one from 2003 to 2010. Historically, most of the outflows of M\&As have been done by developed countries (DCs). However, the proportion of M\&As outflows done by least developed countries (LDCs) has been increasing over time \cite{campi_duenas}. 

In Figure~\ref{fig:total_ma}a we also analyse the contribution of outflows of different groups of countries to total M\&As (see the list of countries in Supplementary Information). The first group, with 37 members, includes countries that belong to the giant strongly connected component in all the years of the sample (i.e. the intersection of members of the giant component in all the years). The second group, with 93 members, includes the set-difference between the union of all countries in the giant component in any year and the members of the first group. 

The first group is mainly composed by developed economies that largely explain the variation in the waves of M\&As. Instead, the second group is mainly composed by developing countries, which explain to a much lesser extent the variations of M\&As, although their participation is much better appreciated in the second wave. A third group (not reported in the figure), with 94 members, represents all those countries that never got strongly connected to the giant component of the network and their investments abroad are practically null.  

This evidence shows that the IMAN is strongly target oriented since most of the investments come from the first group while an important set of countries mainly receive investments, i.e. there are relatively few acquirers and several target markets. Indeed, Figure~\ref{fig:total_ma}b shows that the number of countries participating in the IMAN increases over time, and that countries receiving foreign capitals exceed the quantity of countries investing abroad.

Figure~\ref{fig:total_ma}c shows that the number of links increases over time. However, we observe that the changes in the number of links is less pronounced than the changes in the total volume of M\&As. For instance, between 1995 and 2000, the total volume of M\&As increased around six times, while the number of links only increased 62\% in the same period. This implies that during the growing part of the waves, the IMAN is governed by an intensification of flows (intensive margin) rather than by the creation of new links (extensive margin). An interesting observation is that, in the second wave, the participation of the second group in the number of total links increases, which is also reflected in a higher, but still low, volume of M\&As performed by the group. 

Even though the number of links tends to grow over time, the density is very low and it ranges between 4\% and 5\%, see Figure~\ref{fig:total_ma}d. Density reacts very little to changes in the number of links: in 2007, the year with the highest amount of links, the density is not the highest. This means that during boom periods links grow faster than the squared of the number of participating countries and, therefore, the binary network might be better characterized for having a stable density. In other words, new nodes contribute with very few links with respect to the possible ones. 

Table~\ref{tb:sum_st} presents some additional descriptive statistics for different years. We include statistics for: 1995, the first year of the time series, 2000, the peak of the first wave, 2003, the turning point of the waves, 2007, the peak of the second wave, and 2010, the last year of the time series. 
\begin{table}[h]
  \renewcommand{\arraystretch}{1.1}
  \caption{Summary statistics of the IMAN and its giant strongly connected component.}\label{tb:sum_st}
  \begin{center}
    \begin{small}
      \centering
      \begin{tabular}{l ccccc}
 & 1995 & 2000 & 2003 & 2007 & 2010 \\
        \toprule
        \multicolumn{6}{l}{\textit{International M\&As Network}} \\
        \hline
Countries (No.) & 104 & 129 & 132 & 157 & 149 \\
Mergers (No.) & 95 & 117 & 116 & 148 & 144 \\
Acquirers (No.) & 63 & 87 & 88 & 97 & 96 \\
Links (No.) & 522 & 840 & 715 & 1,159 & 990 \\
Density (\%) & 4.9 & 5.1 & 4.1 & 4.7 & 4.5 \\
Share of reciprocated links (\%) & 18.8 & 20.1 & 19.4 & 21.1 & 21.2 \\
        \hline
Targets making up to 50\% of M\&As & 3 & 2 & 5 & 5 & 6 \\
Targets making up to 90\% of M\&As & 19 & 16 & 29 & 28 & 31 \\
Acquirers making up to 50\% of M\&As & 2 & 2 & 2 & 5 & 6 \\
Acquirers making up to 90\% of M\&As & 12 & 13 & 18 & 24 & 24 \\
Flows making up to 50\% of M\&As & 12 & 9 & 20 & 31 & 34 \\
Flows making up to 90\% of M\&As & 109 & 103 & 161 & 243 & 214 \\
        \hline
        \midrule
        \multicolumn{6}{l}{\textit{Giant Strongly Connected Component}} \\
        \hline
Countries (No.) & 53 & 70 & 68 & 86 & 84 \\
Size Giant Component (\%) & 51.0 & 54.3 & 51.5 & 54.8 & 56.4 \\
Share of Total M\&As (\%) & 95.1 & 99.2 & 94.6 & 98.0 & 94.7 \\
Density (\%) & 15.1 & 14.7 & 12.7 & 13.3 & 12.4 \\
Share of reciprocated links (\%) & 23.6 & 23.8 & 24.0 & 25.2 & 24.3 \\
        \bottomrule
      \end{tabular}
    \end{small}
  \end{center}
\end{table}

The proportion of reciprocated links is around 20\%. The number of acquirers/targets making up a high percentage (50\% or 90\%, respectively) to/from targets/acquires is very concentrated in a few countries, although it slightly increases over time, which may indicate a spread of acquirers to new markets. In addition, the number of flows making up to 50\% and 90\% of M\&As is very concentrated. 

Table~\ref{tb:sum_st} also shows the network statistics of the giant strongly connected component computed every year. We observe that the IMAN is characterized by a giant component and many weakly connected components. The size of the giant component, as a proportion of the number of countries in the network in a given year, reaches up to 56\%, and includes around 95\% of total M\&As. These statistics differ greatly from the corresponding statistics derived for the ITN. Probably, the most relevant difference between the ITN and the IMAN is the level of link reciprocity. In contrast to M\&As relations, which are mainly unilateral, trade relations are typically reciprocal, leading to a higher density and full connectivity \cite{garlaschelli2004patterns, fagiolo2009pre}. However, both the IMAN and the ITN are characterized by highly unbalanced bilateral flows. In the ITN, developed countries have more balanced trade relationships than developing countries \cite{duenas2014global}.

The number of targets and acquirers per country can be obtained from the network statistics $ND_{out}$ and $ND_{in}$, respectively (defined in Methods). There are relevant differences between the distribution of these statistics given the different roles of acquirer and target nodes in the IMAN, and also they present some over-time variations. In the IMAN, only a few countries invest in a high number of targets, which implies that most countries attract capitals from a limited number of countries. This represents a higher concentration of zero out-degrees than zero in-degrees. Another feature is that both distributions are multimodal. In particular, the density of $ND_{out}$ shows many maximums at high degree levels, which also implies that a few countries invest in many targets. We test the stability of these distributions using the two-sample Kolmogorov-Smirnov test (see the results in Supplementary Information). We conclude that there are remarkable differences between in- and out-degree profiles along time.

Agreements among nations have played an important role in the topology of different international networks. In the case of ITN, the proliferation of preferential trade agreements might have lead countries to form intense trade triangles in their neighbourhoods\cite{fagiolo_clus, fagiolo2016oxford}. Given that there is also a proliferation of international investment agreements, we analyse clustering coefficients in order to see if we also observe tight triangles in the case of the IMAN and which type of countries have tighter triplet-like relations. Since the IMAN is very sparse and strongly target oriented, we can expect several types of triplets to coexist in different proportions. For instance, triplets involving two linked developed economies investing in a common target are more likely to be observed than triplets in which there are two linked developing countries.

We observe that the binary and weighted clustering patterns of the IMAN reveal a markedly heterogeneity along time: there are neither well established nor persistent hubs in the network (see the top-5 positions for all directed clustering types in Supplementary Information). For the binary clustering coefficients, the most clustered countries are those known for having favourable legal and fiscal frameworks to attract FDIs --for example, Puerto Rico, which has an established policy to offer huge tax brakes to US-investors-- and tax havens. In the case of weighted clustering, tax havens are less common, while there are mainly high-income economies and emerging market economies. These results shed light on an important aspect of M\&As. Indeed, it is well known that a relevant number of M\&As moves through tax havens and that also other countries get involved either because they are emerging economies or because their main industries are related to commodities of great value in international markets, such as mining and oil.

\subsubsection*{Correlation patterns}

In order to deeply understand the architecture of the IMAN, we compute the correlations among network statistics and their evolution over time. Our aim is to see whether these correlations change during the boom periods, which might indicate different architectures of the IMAN during the waves of M\&As.

First, we analyse whether the total country's M\&As (i.e, node strength) is positively correlated with the country's number of partners (e.g., node degree). Indeed, the top rankings of countries according to node degree and strength are very similar and composed to a great extent by high-income economies (e.g., USA, UK, Germany, Canada, etc.) and by a few emergent countries (e.g., China, Brazil, India) (see node degree/strength rankings in Supplementary Information). This suggests the existence of an over time positive correlation between the two distributions. 

\begin{figure}[!h]
\centering%
\includegraphics[width=\textwidth]{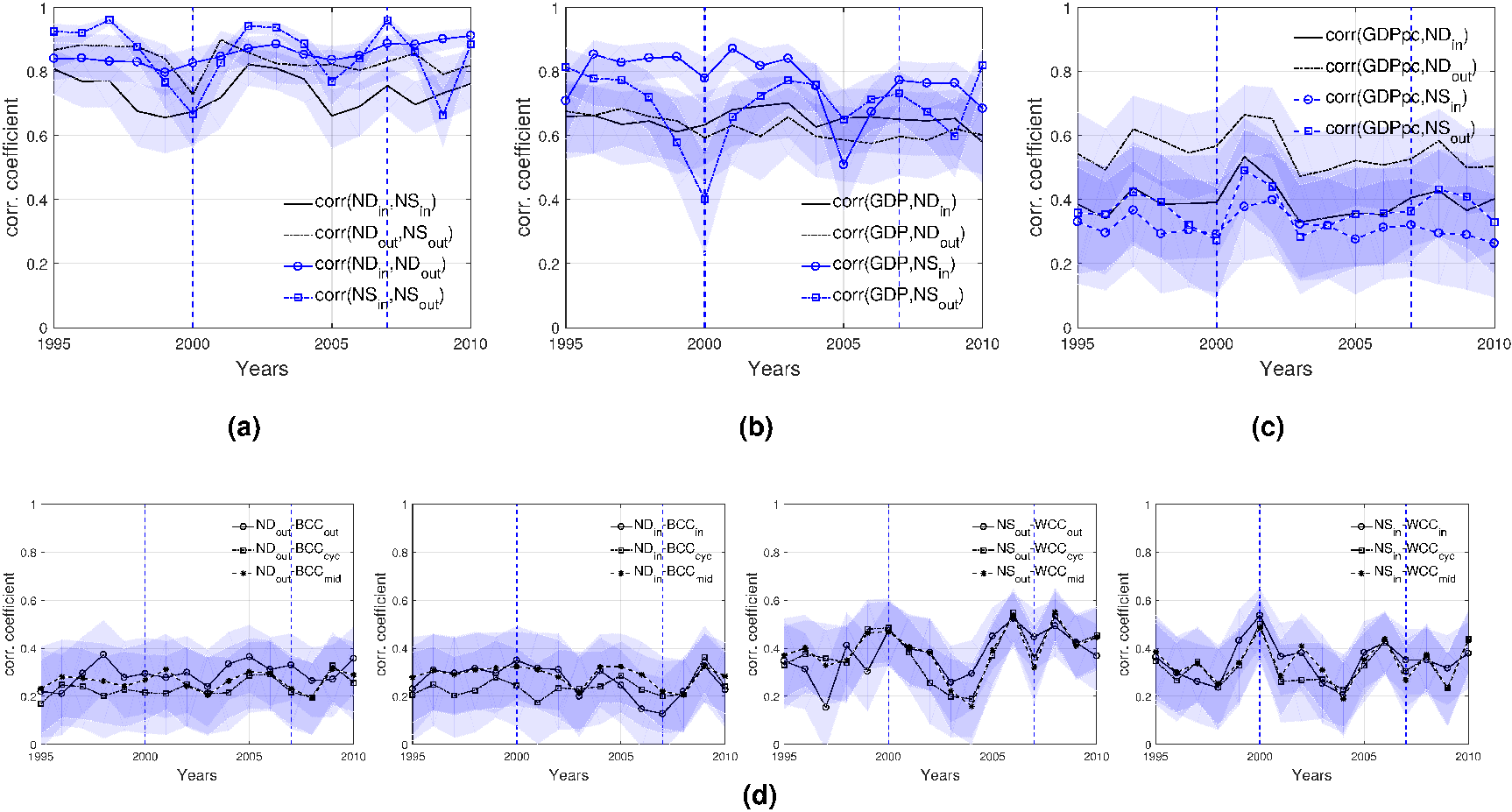}
 \caption{\textbf{Correlation patterns.} \textbf{(a)} Node degree and node strength. \textbf{(b)} Node degree/strength with country size. \textbf{(c)} Node degree/strength with country income. \textbf{(d)} Node degree/strength and binary/weighted clustering coefficients. \textbf{Notes:} Vertical lines indicate the peaks of the two waves. Shaded areas correspond to the 95\% confidence intervals.} \label{fig:cor_ndns}
\end{figure}

Figure~\ref{fig:cor_ndns}a plots the correlation between node degree and node strength for all years. The evidence confirms that countries with more target and acquirer partners have more outflow and inflow investment volumes, correspondingly. Additionally, the within in- and out-statistic correlations are also positive and significant, which indicates a positive relation among the number of targets and acquirers, and also among the amount of outflows and inflows investment volumes. 

We investigate how node degree and node strength relate with size and the income level of countries. Figure~\ref{fig:cor_ndns}b and \ref{fig:cor_ndns}c show the correlations between node degree/strength with countries' GDP and GDP per capita (GDPpc), which are frequently used as determinants of M\&As. We observe that these correlations are generally positive and significant, even if they vary much over time. The correlation is especially higher in the case of the GDP, which might imply that variations of M\&As across countries could be more closely related to their size rather than to their income. Similarly, these correlations are positive high in the ITN and in the  international financial network (IFN) \cite{schiavo_etal_2010, abbate}. 

Finally, we want to test to what extent the IMAN exhibits a structure where countries that are more evolved in M\&As are more intensively clustered. We study the correlations between pairs of node degree/strength and binary/weighted clustering coefficients (Figure~\ref{fig:cor_ndns}d). We observe weak but significant correlations. It is interesting that even if the correlations fluctuate around a certain level, they do not seem to react in a particular way to the waves of M\&As, and those correlations related to the binary representation change less across time than their weighted counterparts. In fact, the correlations for the same weighted quantities are stronger in some years. The reason behind these weak correlations is related to the fact that the binary and weighted clustering patterns change over time.

An interesting pattern emerges from the correlations between: $(ND_{in/out},BCC_{cyc})$, $(ND_{in/out},BCC_{mid})$, $(NS_{in/out},WCC_{cyc})$, and $(NS_{in/out},WCC_{mid})$. Since these types of clustering patterns are characterized by triads in which the node sends and receives foreign investments, it looks like there is a sort of reinforcing mechanism in the motifs of the IMAN. Therefore, an important extent of the volumes go and return to their places of origin, or end up in the acquirer. Thus, a third country gets involved in the looping mechanism, increasing its clustering levels. 

Interesting differences between the IMAN and the ITN can be derived from the clustering analysis. In the ITN, the most clustered countries are those that are high-income economies, trade higher volumes, and have more trade partners. In the IMAN, most clustered countries are not always the richest, nor the same that do or receive the highest volumes or flows of investments. 

\subsection*{The international M\&A network in time}

We have observed that investment decisions are very limited and selective. Given that we have monthly data, we can study if this is related with an annual aggregated behaviour and also how time aggregation creates this network structure. Since we are interested in studying if the network changes at different resolution levels, we define the network in specific time-windows that can range from one month to several months. We focus on network properties of the monthly-cumulated IMAN (more details in the Methods section). 

\begin{figure}[h!]
\centering%
\includegraphics[width=\textwidth]{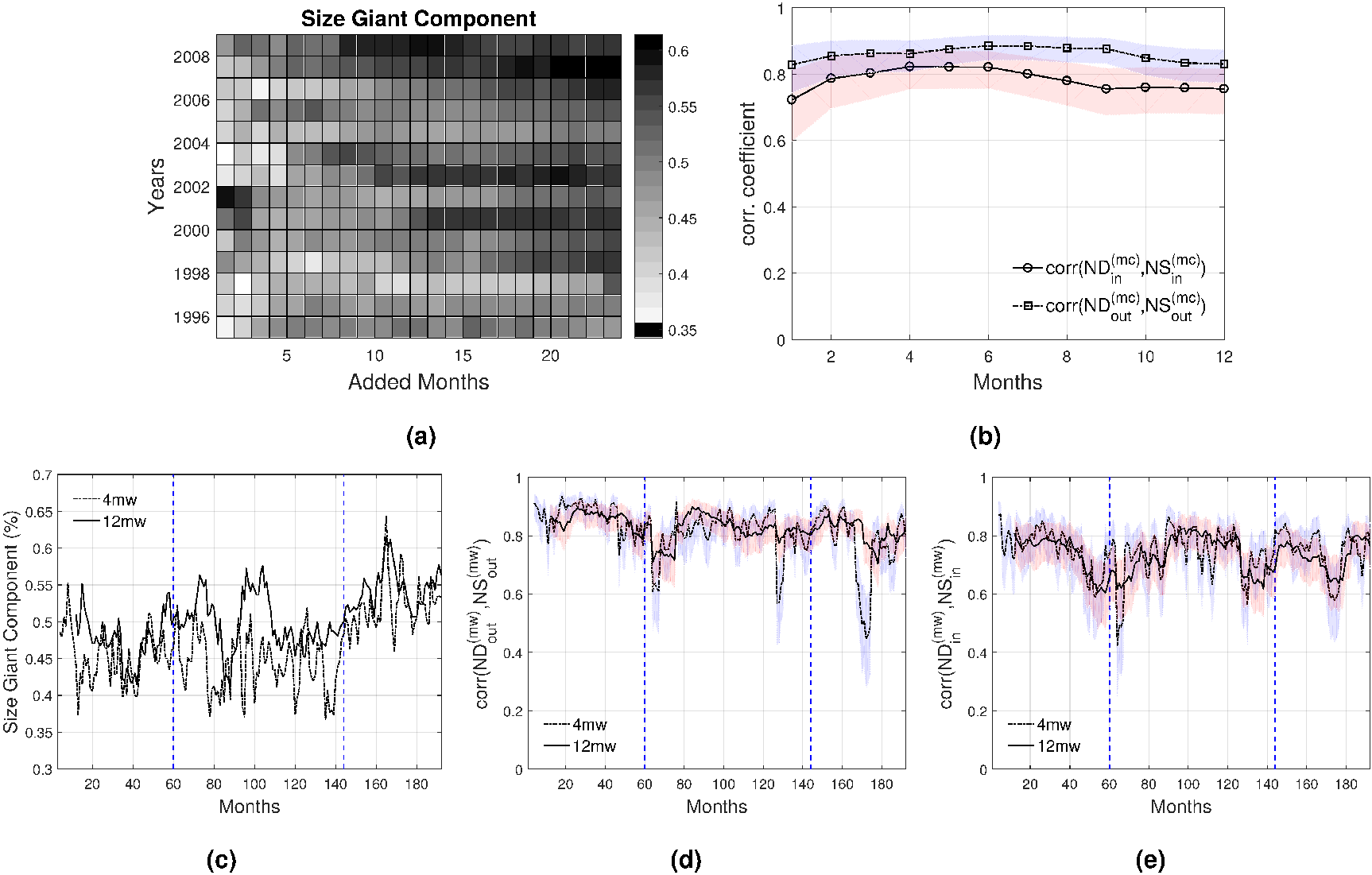}
 \caption{\textbf{Network properties of the monthly-cumulated IMAN representation for fixed (top) and moving windows (bottom).} \textbf{(a)} Size of the giant component relative to the number of countries. \textbf{(b)} Correlations between outward and inward statistics for 2007. \textbf{(c)} Size of the giant component. \textbf{(d)} Correlation between node out-degree and out-strength. \textbf{(e)} Correlation between node in-degree and in-strength. \textbf{Notes:} Vertical lines indicate the peaks of the two waves. Shaded areas correspond to the 95\% confidence intervals.}\label{fig:giant_time}
\end{figure}

Figure~\ref{fig:giant_time}a shows the size of the giant component for the monthly-cumulated representation for different years. We observe that the giant component is already evident when aggregating over one month, as its size is very high (more than 35\%), compared with the maximum size observed for yearly data (just above 50\%). The 50\% level is reached quickly, in between 3 and 4 months, and for longer time-windows, the giant component grows very little and very irregularly with some fluctuations. Moreover, if we aggregate up to 24 months, the size increases, but does not reach 100\%. In other words, the network never gets strongly connected. 

The fact that the giant component grows modestly with the aggregation of links indicates that an intensive margin of investments governs over the extension to new partners. This means that existing relationships benefit from higher investments and that there is little expansion to new markets. Moreover, the isolated and weakly connected nodes hardly get strongly connected with the aggregation of months.

Figure~\ref{fig:giant_time}b shows the correlations between pairs of node degree and node strength for the monthly-cumulated representation of the IMAN in 2007, selected as a benchmark given the highest cross-border volumes of M\&As in this year. These correlations are stationary and already strong in the smallest time-window considered (one month). This evidence suggests that the network structure does not react much to the increase in the number of links. Since the giant component is present in every window, and it is the only component with more than one member, we can say that it fully characterizes the patterns of the IMAN. Therefore, weakly connected countries play a marginal role and do not contribute to a large extent to the structure of the network.

We might expect the architecture of the network to react differently to the worldwide booms of M\&As. In order to explore this, we use a moving-window network approach. Thus, given an initial month, $t_0$ and a fixed $\Delta t$, we consider moving windows of one step (see Methods section). Figure~\ref{fig:giant_time}c shows the evolution of the size of the giant component. We use two different time-window sizes, $\Delta t=$4 and $\Delta t=$12, in order to observe possible differences in the short- and medium-term. Also, we have included two vertical lines indicating the beginning of the years 2000 and 2007, in which we observe the peaks of the two waves. 

We observe that the size of the giant component fluctuates around a certain average level and that it does not change during the first wave. However, in accordance with Table~\ref{tb:sum_st}, the giant component seems to be enlarged after the 2007 boom. Data for more recent years would allow us to study if this change in the size of the giant component also implies a change in how it reacts to the waves. However, we could argue that this enlargement of the size of the giant component is more related with the increasing integration of LDCs (or countries in group 2) to the IMAN, as we observed on Figure~\ref{fig:total_ma}. 

Figure~\ref{fig:giant_time}d and \ref{fig:giant_time}e show the evolution of correlations between pairs of node degree and node strength. While these correlations fluctuate, on average, there are no major differences between different time-windows. 

We can conclude that time aggregation do not change the network properties and also that these features are present in short time-windows, including the periods of booms and decline. Thus, the characteristics of the network are quite robust over time. Similarly, Fagiolo et al. \cite{fagiolo2010evolution} show that the structural properties of the ITN display a remarkable stability and claimed that recent trade integration has not had a significant impact on the structure of the ITN.

\subsection*{The international M\&A network in space}

In this section, we study how the geographical space affects the topology of the IMAN. Geographical distances might be able to capture bilateral characteristics of countries and to shed light on the barriers conditioning countries interactions. For the spatial analysis we use yearly data because, as we saw in the temporal analysis, the structure of the network can be completely appreciated in this time lapse, and because the analysis of most international networks are done with yearly data. 

The idea consists in using the geographical distances between countries reporting a transaction to build a sequence of sub-networks. Links in each sub-network are expected to share similar distances and, for this reason, expected to be related to similar barriers to invest. Then, we analyse how the topological properties of the sub-networks change as distance increases. The shape of the distribution of distances is left-skewed, bi-modal with a higher peak at farther distances, and very stable (see Supplementary Information).

We compute the deciles of the distribution of geographical distances and use them to split the IMAN into two different families of sub-networks. The \textit{restricted representation}, consists of sub-graphs derived by keeping the links associated to geographical distances between the limits of a given decile in a given year. The \textit{cumulated representation}, consists of sub-graphs derived by keeping the links with geographical distances below the upper limit of a given decile. These network families are complementary. The cumulated framework gives us a picture of the IMAN for links constrained between zero and a given threshold, while the restricted framework gives us a picture of the IMAN constrained to a certain range (more details in the Methods section). 

\begin{table}[h!]
  \caption{Summary of the network statistics for the restricted representation (by deciles intervals).}\label{tb:sum_restricted}
\begin{center}
\begin{small}
\centering
\begin{tabular}{l cccccccccc}
\toprule 
 & 1st  & 2nd  & 3rd  & 4th  & 5th  & 6th  & 7th  & 8th  & 9th  & 10th  \\
\scriptsize Dist.(Km.) & \scriptsize 0-764 & \scriptsize 765-1,342 & \scriptsize 1,343-1,923 & \scriptsize 1,924-2,852 & \scriptsize 2,853-4,653 & \scriptsize 4,654-6,280 & \scriptsize 6,281-7,965 & \scriptsize 7,966-9,589 & \scriptsize 9,590-12,092 & \scriptsize 12,093-19,147 \\
\midrule
& \multicolumn{10}{c}{Share of total M\&As (\%)} \\
\hline
1995 & 25 & 10 & 2 & 3 & 2 & 27 & 8 & 3 & 5 & 15\\ 
2000 & 37 & 5 & 6 & 5 & 2 & 22 & 12 & 4 & 4 & 3\\ 
2003 & 17 & 6 & 9 & 5 & 6 & 31 & 8 & 4 & 7 & 8\\ 
2007 & 25 & 9 & 7 & 5 & 3 & 23 & 8 & 4 & 8 & 8\\
2010 & 16 & 5 & 4 & 6 & 14 & 18 & 10 & 9 & 8 & 10\\ 
\midrule
& \multicolumn{10}{c}{Averages of Link Weights} \\
\hline
1995 & 861 & 315 & 79 & 90 & 67 & 887 & 262 & 115 & 166 & 507 \\ 
2000 & 5,147 & 639 & 913 & 676 & 246 & 3,023 & 1,733 & 613 & 605 & 379 \\ 
2003 & 655 & 237 & 334 & 194 & 227 & 1,226 & 316 & 146 & 272 & 313 \\ 
2007 & 2,848 & 1,002 & 825 & 562 & 290 & 2,698 & 921 & 485 & 888 & 948 \\ 
2010 & 899 & 251 & 252 & 344 & 775 & 1,024 & 549 & 491 & 468 & 575 \\ 
\midrule
& \multicolumn{10}{c}{Averages of Node Degrees} \\
\hline
1995 & 3.4 & 2.8 & 2.7 & 1.9 & 2.1 & 3.3 & 2.5 & 2.3 & 3.3 & 3.0 \\ 
2000 & 3.9 & 3.2 & 3.6 & 3.4 & 2.7 & 3.0 & 3.2 & 3.2 & 3.5 & 3.5 \\ 
2003 & 3.5 & 3.1 & 3.0 & 2.1 & 2.6 & 3.2 & 2.8 & 2.9 & 3.1 & 3.4 \\ 
2007 & 3.8 & 4.0 & 3.1 & 3.0 & 2.9 & 3.8 & 2.9 & 3.4 & 3.8 & 4.0 \\ 
2010 & 3.2 & 3.1 & 2.9 & 2.9 & 2.8 & 4.0 & 2.8 & 3.6 & 3.5 & 3.9 \\ 
\midrule
& \multicolumn{10}{c}{Averages of Out-Binary Clustering Coefficients} \\
\hline
1995 & 0.14 & 0.03 & 0.01 & 0.00 & 0.01 & 0.00 & 0.00 & 0.00 & 0.01 & 0.00\\ 
2000 & 0.23 & 0.02 & 0.04 & 0.00 & 0.01 & 0.01 & 0.00 & 0.00 & 0.00 & 0.00\\ 
2003 & 0.21 & 0.04 & 0.04 & 0.01 & 0.00 & 0.00 & 0.00 & 0.00 & 0.00 & 0.00\\ 
2007 & 0.13 & 0.01 & 0.02 & 0.02 & 0.03 & 0.01 & 0.00 & 0.00 & 0.01 & 0.00\\ 
2010 & 0.12 & 0.01 & 0.01 & 0.02 & 0.01 & 0.00 & 0.00 & 0.00 & 0.00 & 0.00\\ 
\midrule
& \multicolumn{10}{c}{Averages of In-Binary Clustering Coefficients} \\
\hline
1995 & 0.17 & 0.03 & 0.02 & 0.00 & 0.01 & 0.00 & 0.00 & 0.00 & 0.01 & 0.00 \\ 
2000 & 0.25 & 0.04 & 0.04 & 0.00 & 0.01 & 0.01 & 0.00 & 0.00 & 0.01 & 0.00 \\ 
2003 & 0.21 & 0.05 & 0.04 & 0.01 & 0.00 & 0.00 & 0.00 & 0.00 & 0.00 & 0.01 \\ 
2007 & 0.15 & 0.01 & 0.02 & 0.02 & 0.03 & 0.01 & 0.00 & 0.00 & 0.02 & 0.00 \\ 
2010 & 0.14 & 0.01 & 0.02 & 0.02 & 0.00 & 0.00 & 0.01 & 0.00 & 0.00 & 0.00 \\ 
\bottomrule
\end{tabular}
\end{small}
\end{center}
\end{table}

Table~\ref{tb:sum_restricted} shows the summary of the network statistics in the restricted representation. We observe that cross-border investments distribute mostly in relatively short and intermediate distances, as the higher shares are observed for the first and sixth deciles. The average of link weights is comparatively higher for links with very short and medium geographical distances (first and sixth deciles, respectively). Conversely, the average of node degrees has small variations across distance deciles, although in most years the higher values are observed at the shortest and longest distances (first and tenth deciles). This means that countries have many relationships with countries in their own geographical neighbourhood but also with countries located relatively far from them. 

Long distance interactions do not play an important role for the binary clustering coefficients. Indeed, they are equal to zero for most decile intervals, except for the first ones. Hence, the probability that any pair of partners of a node are themselves partners, subject to that all countries are separated by long geographical distances, is practically null.

Table~\ref{tb:sum_cumulative} shows the averages of the network statistics for the cumulated network representation. It exists a decreasing and non-linear relationship between the average of link-weights and the distance, as can be observed in the changes of the averages of link weights due to the addition of a decile. The averages of link weight decrease from the first until the fifth decile, they increase in the sixth decile, and decrease afterwards. This behaviour is observed for all years, except in the year 2010, in which the increase is observed in both the fifth and sixth deciles. The decrease of the averages of link weights is rather moderate compared with the dramatic decrease observed in the case of trade flows\cite{abbate}. 

\begin{table}[h!]
 \caption{Summary of the network statistics for the cumulative representation (by deciles intervals).}\label{tb:sum_cumulative}
\begin{center}
\begin{small}
\centering
\begin{tabular}{l cccccccccc}
\toprule 
 & 1st  & 1st-2nd  & 1st-3rd  & 1st-4th  & 1st-5th  & 1st-6th  & 1st-7th  & 1st-8th  & 1st-9th  & 1st-10th  \\
\scriptsize Dist.(Km.) & \scriptsize 0-764 & \scriptsize 0-1,342 & \scriptsize 0-1,923 & \scriptsize 0-2,852 & \scriptsize 0-4,653 & \scriptsize 0-6,280 & \scriptsize 0-7,965 & \scriptsize 0-9,589 & \scriptsize 0-12,092 & \scriptsize 0-19,147 \\
\midrule
& \multicolumn{10}{c}{Averages of Link Weights} \\
\hline
1995 & 861 & 585 & 418 & 335 & 281 & 385 & 367 & 336 & 317 & 335\\ 
2000 & 5,147 & 2,955 & 2,257 & 1,871 & 1,539 & 1,791 & 1,780 & 1,636 & 1,520 & 1,407\\ 
2003 & 655 & 445 & 408 & 355 & 329 & 478 & 455 & 416 & 400 & 391\\ 
2007 & 2,848 & 1,933 & 1,564 & 1,313 & 1,109 & 1,372 & 1,307 & 1,204 & 1,169 & 1,147\\ 
2010 & 899 & 572 & 466 & 435 & 502 & 589 & 582 & 572 & 559 & 562\\ 
\midrule
& \multicolumn{10}{c}{Change in Averages of Link Weights (\%)} \\
\hline
1995 &  & -32.0 & -28.7 & -19.8 & -16.1 & 37.0 & -4.6 & -8.5 & -5.7 & 6.0\\ 
2000 &  & -42.6 & -23.6 & -17.1 & -17.8 & 16.4 & -0.6 & -8.1 & -7.0 & -7.5\\ 
2003 &  & -32.1 & -8.2 & -13.1 & -7.2 & 45.2 & -4.8 & -8.5 & -3.9 & -2.2\\ 
2007 &  & -32.1 & -19.1 & -16.0 & -15.5 & 23.7 & -4.7 & -7.9 & -2.9 & -1.9\\ 
2010 &  & -36.4 & -18.5 & -6.6 & 15.4 & 17.3 & -1.2 & -1.8 & -2.2 & 0.5\\ 
\midrule
& \multicolumn{10}{c}{Averages of Binary Out-Clustering Coefficients} \\
\hline
1995 & 0.14 & 0.17 & 0.15 & 0.13 & 0.15 & 0.14 & 0.16 & 0.15 & 0.18 & 0.19\\ 
2000 & 0.23 & 0.12 & 0.16 & 0.17 & 0.19 & 0.17 & 0.16 & 0.15 & 0.15 & 0.17\\ 
2003 & 0.21 & 0.12 & 0.13 & 0.15 & 0.14 & 0.14 & 0.15 & 0.14 & 0.15 & 0.16\\ 
2007 & 0.13 & 0.13 & 0.15 & 0.18 & 0.18 & 0.18 & 0.18 & 0.17 & 0.17 & 0.19\\ 
2010 & 0.12 & 0.10 & 0.15 & 0.15 & 0.14 & 0.15 & 0.15 & 0.15 & 0.15 & 0.16\\ 
\hline
& \multicolumn{10}{c}{Averages of Binary In-Clustering Coefficients} \\
\hline
1995 & 0.17 & 0.25 & 0.30 & 0.33 & 0.32 & 0.33 & 0.33 & 0.35 & 0.39 & 0.41\\ 
2000 & 0.25 & 0.23 & 0.32 & 0.36 & 0.39 & 0.35 & 0.38 & 0.38 & 0.39 & 0.40\\ 
2003 & 0.18 & 0.18 & 0.25 & 0.29 & 0.31 & 0.32 & 0.33 & 0.37 & 0.38 & 0.40\\ 
2007 & 0.16 & 0.23 & 0.24 & 0.32 & 0.36 & 0.38 & 0.40 & 0.43 & 0.43 & 0.46\\ 
2010 & 0.14 & 0.19 & 0.26 & 0.32 & 0.29 & 0.29 & 0.31 & 0.34 & 0.34 & 0.36\\ 
\bottomrule
\end{tabular}
\end{small}
\end{center}
\end{table}

The average of $BCC_{out}^{(c)}$ increases very slowly with distance. This implies that, on average, the probability that two target partners are linked is low, i.e. between 10\% and 23\%, regardless the geographical distance between them. Instead, the average of $BCC_{in}^{(c)}$ increases with distance, implying that the probability that two acquirer partners are linked increases as more distant links are attached to the network. It is worth noticing that the population of target countries is greater than the population of acquirer countries, which agrees with the fact that the average $BCC_{out}^{(c)}$ is lower than the average of $BCC_{in}^{(c)}$, in most deciles.
 
\begin{figure}[h!]
\centering%
\includegraphics[width=\textwidth]{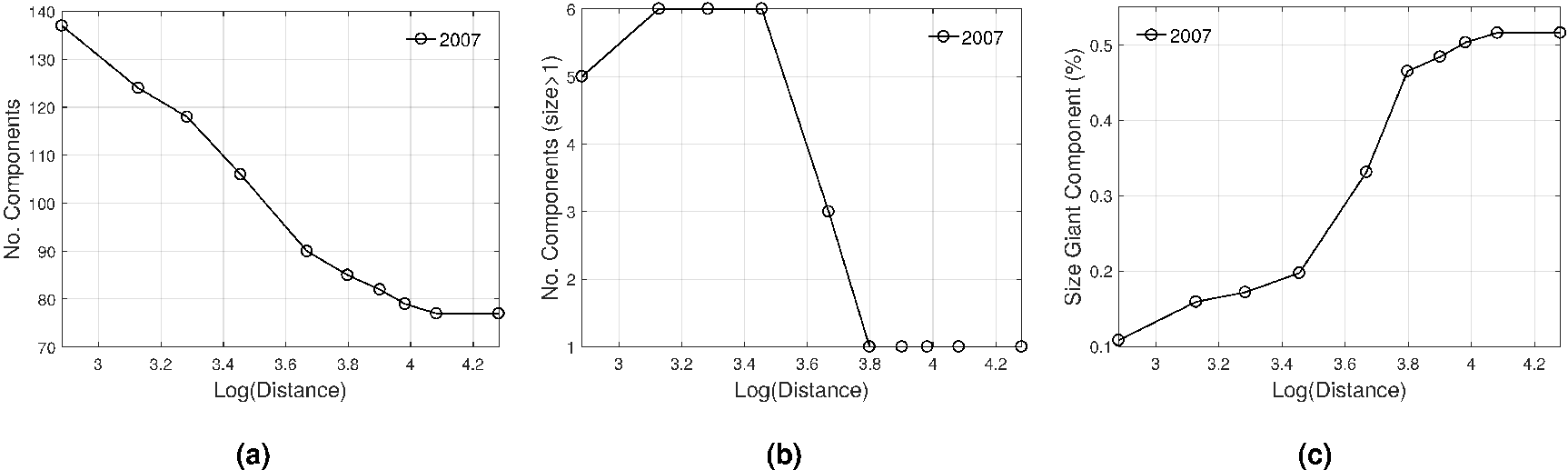}
\caption{\textbf{Network statistics of the cumulated representation in 2007.} \textbf{(a)} Number of components. \textbf{(b)} Number of components of size greater than one. \textbf{(c)} Size of the giant component relative to the number of countries.}\label{fig:components}
\end{figure}

Figure~\ref{fig:components}a displays the number of components and the size of the giant component for the cumulated IMAN. The connectivity changes as links with longer geographical distances join the network. The number of connected components decreases with distance from around 140 to less than 80 components, which represents a huge number of components. But, in Figure~\ref{fig:components}b, we observe that if we exclude isolated nodes, the number of components falls until 5 in the first decile, and until only one component from the sixth decile onwards. 

Therefore, only a few components are able to grow as more distant links are allowed. Around the sixth decile, all those components get interconnected giving rise to the giant component. The rest of the weakly connected components are isolated countries, which are targets of acquirers that belong to the giant component. Figure~\ref{fig:components}c shows the growing pattern of the size of the giant component, which grows approximately five times as more distant links are added, until it reaches almost 50\% of total countries.

\subsubsection*{Spatial correlation patterns}

In this section we focus on correlation patterns between networks statistics for the cumulated representation to shed light on the role, if any, of geographical distance in shaping the IMAN structure. We aim at understanding to what an extent spatial connectivity affects local and higher order network properties, including clustering and assortativity.

\begin{figure}[h!]
\centering%
\includegraphics[width=\textwidth]{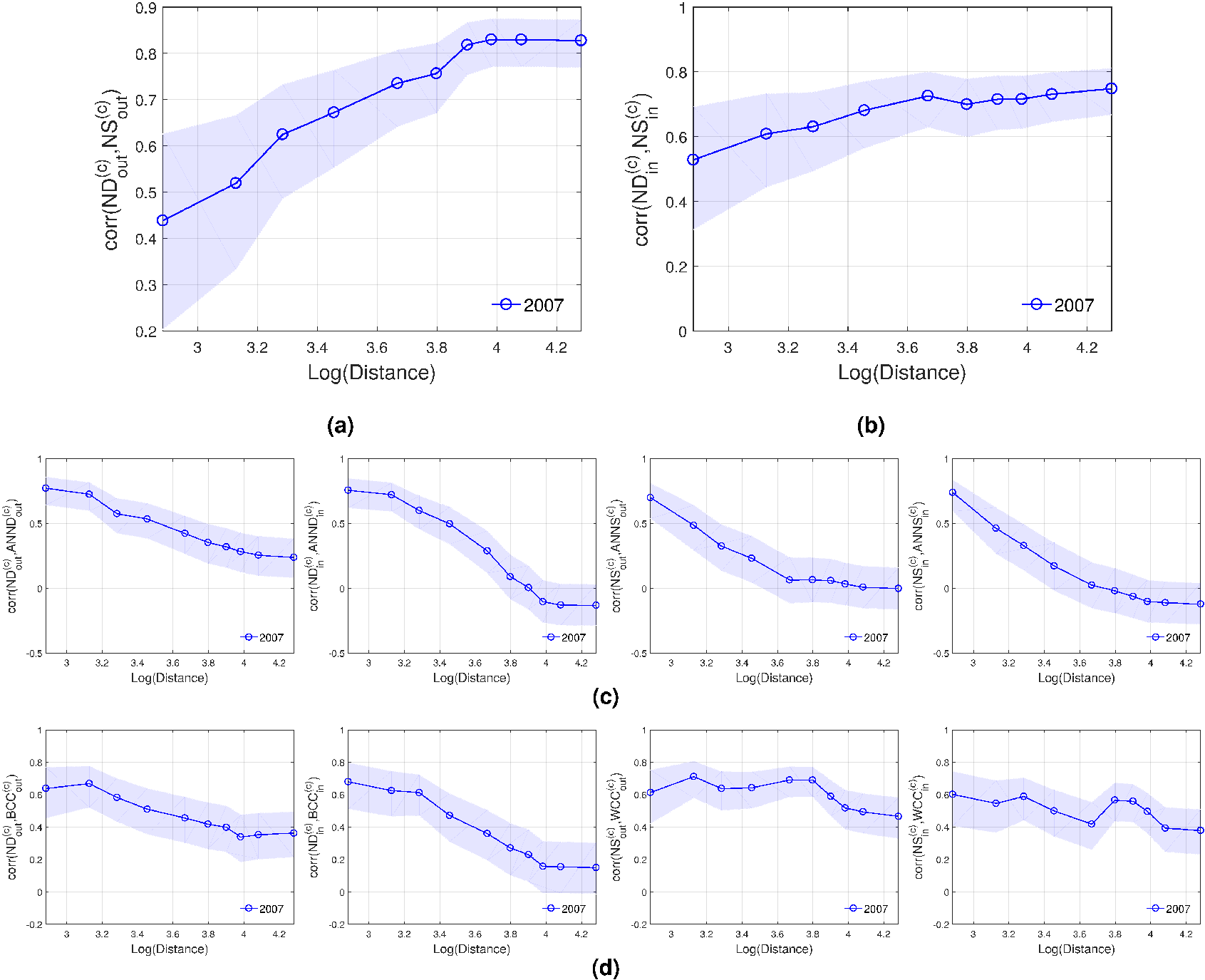}
 \caption{\textbf{Spatial correlation patterns of the cumulated representation}. \textbf{(a)} Node out-degree and out-strength correlations. \textbf{(b)} Node in-degree and in-strength correlations. \textbf{(c)} Assortative patterns. \textbf{(d)} Node degree/strength correlations with clustering coefficients. Shaded areas correspond to the 95\% confidence intervals.}\label{fig:qs_corrs}
\end{figure}

Figure~\ref{fig:qs_corrs}a and \ref{fig:qs_corrs}b show the correlation between node degree and strength, for the cumulated distance networks in 2007. We observe positive correlations for all geographical distances. Hence, countries with many acquirer partners are also receiving high levels of investments, and similarly, countries with many target partners are making high levels of investments. This evidence might be against the hypothesis that geographical distances represent barriers to invest. We observe that the geographical extension of markets and the intensity of the transactions are positively correlated, and that the correlation gets stronger when long-distant pairs are considered. 

Figure~\ref{fig:qs_corrs}c shows the correlation coefficients between (in/out) node degree/strength and average nearest-neighbor degree/strength. We find that both binary and weighted cumulated representations exhibit a very assortative pattern at short distances with the correlation becoming quickly non-significant as the limit of the geographical distance increases (except for the binary outflow case). The assortative pattern survives until the third and fifth deciles. This result could be explained in part by the existence of international investment agreements and bilateral investment treaties that are quite diffused at the regional level. On the other hand, assortativity might be broken by the wealth distribution in the world: high volumes of capital are concentrated in a few countries, giving them the possibility to invest in several others, which do not necessarily have many cross-border relationships. These results can be compared with those obtained for the ITN, which also exhibits an assortative pattern when links are restricted to short distances. In contrast, the ITN is dissasortative when links are not restricted to geographical distances \cite{abbate}. 

The correlation coefficients between node degree/strength and clustering coefficients are positive and significant for most distance deciles (Figure~\ref{fig:qs_corrs}d). In the binary representation this means that if one considers only links below a certain geographical distance threshold, it turns out that countries that hold more target/acquirer partners are more clustered than those with a few target/acquirer partners. However, this pattern fades away when the threshold gets higher, since the computed correlations decrease as more distant links join the network. In the weighted representation, if links are constrained to a certain distance threshold, countries with high intensity of M\&As relationships are typically involved in intense interconnected triplets. In contrast to the binary counterpart, the computed correlations in the weighted network remain roughly constant across different distance deciles.

Overall, the spatial analysis shows that the architecture of the restricted IMAN at different geographical distance deciles exhibit different topologies, which in turn may be different from those observed in the whole network. On the other side, the structure of the cumulated sub-networks display properties that move towards the one observed for the whole network.

\subsection*{Discussion}

In this paper we developed a comprehensive study of the international M\&As network using a complex network approach. Firstly, we studied the topological architecture of the network. Secondly, we focused on the dynamics of the network statistics over time. Finally, we analysed how the geographical distance affects the structure of the network. 

Given that trade and M\&As are two common strategies of firms to reach a foreign market, we employed network statistics frequently used in the analysis of the ITN, and highlighted the differences and similarities between the IMAN and the ITN. Our analysis contributes to the literature providing new and complementary insights to the understanding of international macroeconomics from a network perspective.

The analysis reveals several interesting features of the international M\&A network. Firstly, cross-border M\&As are performed only by a limited group of countries, which invest in a high number of targets, implying also that most countries attract capitals from a limited number of countries. This makes the IMAN strongly target-oriented with a very low proportion of reciprocated links. There is a strong positive correlation between node degree and node strength, which is explained by the target-oriented feature of the IMAN. Secondly, the configuration of the IMAN in the very short and the long term consists of many weakly connected low-income target countries together with a relatively low number of high-income strongly connected economies. Therefore, the IMAN is well characterized by a giant component and many weakly-connected nodes that only have a unilateral link with the remaining network. Thirdly, clustering patterns are very heterogeneous. There are neither well established nor persistent hubs in the network, and more clustered countries are not necessarily the most connected ones. 

Some of these properties contrast with those of the ITN, which is a highly dense network, with a majority of reciprocated links, a stable giant component corresponding to the whole network, and robust clustering patterns. However, both the IMAN and the ITN are characterized by a very dominant group of rich countries that have more reciprocal interactions and shape, to a great extent, the topological structure of these international networks.

The temporal analysis showed that, despite M\&As have had periods of impressive growth and decline, the topological architecture of the network remains quite invariant over time. We showed that the structure of the network does not depend on any particular time aggregation of monthly data. Indeed, increasing the length of the time-window used to aggregate monthly data does not change the properties of the network, including during periods of boom and decline. Thus, the characteristics of the network are quite robust over time. Similarly, the structural properties of the ITN display a remarkable stationarity and recent trade integration has not had a significant impact on the structure of the ITN \cite{fagiolo2010evolution}.

The spatial analysis revealed the existence of a non-linear relationship between M\&As and geographical distance. Cross-border investments distribute mostly in relatively short and intermediate distances. This has two important implications for the architecture of the network: i) distant interactions allow the giant component to get fully connected at medium distances, and ii) short distance interactions generate a very assortative pattern, meaning that well-connected countries tend to interact with other well-connected countries.

Moreover, we showed that link-weights are higher at short and medium geographical distances and that countries have many relationships with countries in their own geographical neighbourhood but also with countries located relatively far from them. The correlation between the geographical extension of M\&As markets and the intensity of the transactions is positive and gets stronger when long-distant investment partners are considered. Therefore, barriers to invest that get stronger with longer geographical distance do not seem to represent a real obstacle to long-distance transactions. 

The non-linear effects of geographical distance are also present in the ITN and other international financial networks, in which a relevant part of the interactions take place in geographical neighbourhoods, but several network patterns need longer distances to emerge. Our results complement recent studies on the geographical properties of international complex networks\cite{abbate, zhang2016}, contributing to the theoretical and empirical debates on the effect of the geographical distance on cross-border M\&As.

Our results might be compared with those presented by Garas et al. \cite{garas2016} in their description of the IFDIN. Both the IMAN and the IFDIN have a low density and low clustering. Conversely, the IFDIN has a negative assortative pattern, while the IMAN has a positive assortativity for links restricted to relative short distances, and no significant pattern for unrestricted distances. However, there are important differences in their approach and the one presented in this paper. First, they use a symmetrized representation, which leads to important implications for the connectivity. Nevertheless, due to the nature of cross-border investments we found that acquires and targets behave very differently. Thus, the direction of the investments certainly plays an important role in understanding the architecture of cross-border investments. Secondly, they use stocks of FDI because they consider that flows are very volatile and therefore harder to model. Instead, we used flows of M\&As and we observed that the main features of the architecture of the IMAN are present both at very short and long time spans. 

Overall, our analysis attempted to provide a network characterization of the IMAN that can contribute to future research on international networks. Borrowing from the broad literature related to the ITN, a possible extension of this research would be to explore the determinants of the architecture of the network following traditional economic modeling\cite{duenas2013modeling} and null network models\cite{fagiolo2013null}. In addition, given the interplay and complementary of trade, migration, financial investment, and M\&As flows, it would be interesting to develop a multilayer network analysis\cite{schiavo_etal_2010, duenas2014global, garas2016}. This could provide new insights on the international relations, and the movement of goods and factors of production that shape globalization.

\section*{Methods}

The data for the analysis are extracted from Worldwide Mergers, Acquisitions, and Alliances Databases SDC Platinum (Thomson Reuters), a collection of financial databases that provide extensive large-scale information on global transactions since 1985 to 2010. Our study covers the period 1995-2010 because these are the years that have complete time series for all countries. 

Most of the recorded transactions (in volume) refer to domestic M\&As-activity (74\%, on average). The domestic links represent around 10\%, indicating that, on average, M\&As tend to be more oriented towards abroad, but the average volume of foreign operations is much lower.

The nominal monthly M\&As inflows and outflows (millions of current USD) are deflated using the Industrial Production Index provided by the US Bureau of Labor Statistics\cite{US} to build a set of directed adjacency matrices from the real M\&As data with rows indicating acquirer countries (i.e. investors) and columns standing for M\&As targets. Our choice to focus on a directed network is driven by the need to keep a clear distinction between who invests and who receives the foreign capital inflows. In fact, our analysis shows that the network is strongly asymmetric: this also suggests that a directed-network analysis is preferred.

Using these data, we can define the IMAN in its weighted and binary representation where nodes are countries linked by capital flows. We distinguish between \textit{acquirers}, countries investing abroad, and \textit{targets}, countries hosting capitals from a foreign country.

The \textbf{\emph{Weighted International M\&A Network}} in a given point of time $t$ is represented by a weighted-directed graph, where the nodes are the $N(t)$ countries and link weights are fully characterized by the $N(t) \times N(t)$ asymmetric matrix $W(t)$, with entries $w_{ij}(t)$, i.e. a flow of M\&As from country $i$ to country $j$.

The \textbf{\emph{ Binary International M\&A Network}} in a given point of time $t$ is represented by a binary-directed graph, where the nodes are the $N(t)$ countries and binary links are fully characterized by the $N(t) \times N(t)$ asymmetric adjacency matrix $A(t)$, with entries $a_{ij}(t)=1$ if and only if $w_{ij}(t)>0$, i.e. M\&As flows from country $i$ to country $j$ are strictly positive.

\subsubsection*{The IMAN in time}

Cross-border M\&As have periods of impressive growth and decline that might affect the architecture of the IMAN. In order to recognize the evolving patterns of the topology of the network, we propose to study sequences of filtered sub-networks, restricted to specific time periods. More precisely, let $Z(\tau) \equiv (z_{ij}(\tau))_{1 \le i,j \le N(\tau)}$ be a temporal network for a specific month $\tau$. Thus, the monthly-cumulated network at time $t$ is defined as:
\begin{equation}
\label{eq:mw_nets}
Z^{(mc)}(t;\Delta t)=\sum_{\tau=t}^{t + \Delta t}Z(\tau);
\end{equation}
where, $\Delta t$ is the length of the time-window, and $Z(\tau)$ can be either $A(\tau)$ or $W(\tau)$. In this way, we aim at analysing the cumulated properties of the IMAN for any time-window $[t,t + \Delta t]$.

\subsubsection*{The IMAN in space}

Geography matters for the interaction between countries. In gravity models, the geographical distance is commonly employed as a proxy of transaction costs, for example, it is frequently considered for the estimation of international trade, migration, and foreign direct investment. In order to understand how the architecture of the IMAN relies on the distance, we propose to analyse two different families of sub-networks capturing links with similar distance ranges. These families are: the \textit{restricted matrices} and the \textit{cumulated matrices}.

Thus, we define the geographical-distance matrix $D$ as; 
\begin{equation}
D=
\begin{cases}
d_{ij}, & \text{if } a_{ij}=1, \\
0, & \text{otherwise};
\end{cases}
\end{equation}
whose generic element $d_{ij}$ is equal to the geographical distance between countries $i$ and $j$, computed using the great-circle formula, which uses latitudes and longitudes of the most important cities/agglomerations (in terms of population) of each country in the pair\cite{CEPII:2011-25}. This matrix is asymmetric as the IMAN has very low levels of reciprocity. See the distribution of distances in Supplementary Material.

Let $Z(t) \equiv (z_{ij}(t))_{1 \le i,j \le N}$ be a generic adjacency matrix (binary or weighted) at time $t$; let introduce $\delta_k$, with $k=1,2,...,10$, the deciles of the distribution of geographical distances ($\delta_0=0$). In each year, the \textit{restricted matrices} are defined as:   
\begin{equation}
Z^{(r)}_k=
\begin{cases}
z_{ij,k}^{(r)}=z_{ij}, & \text{if } \delta_{k-1}< d_{ij} \leq \delta_k, \\
z_{ij,k}^{(r)}=0, & \text{otherwise};
\end{cases}
\end{equation}
and the \textit{cumulated matrices} are defined as:
\begin{equation}
Z^{(c)}_k=
\begin{cases}	
z_{ij,k}^{(c)}=z_{ij}, & \text{if } d_{ij} \leq \delta_{k}, \\
z_{ij,k}^{(c)}=0, & \text{otherwise}.
\end{cases}
\end{equation}

\subsection*{Network Statistics}

We use node statistics that are commonly employed in the international networks of countries\cite{lic03,sebo03,garla2004,fagiolo2009pre}. These statistics allow studying node characteristics in terms of connectivity and clustering. The binary statistics are: node degree ($ND$), average nearest-neighbour degree ($ANND$), and the clustering coefficient ($BCC$). These statistics can be generalized to the weighted network: node strength ($NS$), average nearest-neighbour strength ($ANNS$), and the weighted clustering coefficient ($WCC$). Both sets of statistics are used in their corresponding directed versions. Table~\ref{tb:net_stats} defines the network statistics employed. 
\begin{table}[ht!]
\renewcommand{\arraystretch}{2.2}
\caption{Binary and weighted topological statistics.}\label{tb:net_stats}
\begin{center}
\begin{small}
\begin{tabular}{p{3cm} l l}
\toprule
 Topological Properties & Binary & Weighted \\
\midrule
    {Degrees/Strengths}
    & $ND_{i}^{out}=A_{(i)}\textbf{1}$ & $NS_{i}^{out}=W_{(i)}\textbf{1}$ \\
    & $ND_{i}^{in}=A'_{(i)}\textbf{1}$ & $NS_{i}^{in}=W'_{(i)}\textbf{1}$ \\
    & $ND_{i}^{tot}= ND_{i}^{in}+ND_{i}^{out}$ & $NS_{i}^{tot}= NS_{i}^{in}+NS_{i}^{out}$ \\
\midrule
ANND/ANNS
 & $ANND^{out}_i=\frac{(A+A')_{(i)}(A+A')\textbf{1}}{ND^{out}_i}$ & $ANNS^{out}_i=\frac{(W+W')_{(i)}(A+A')\textbf{1}}{ND^{in}_i}$ \\
& $ANND^{in}_i=\frac{(A+A')_{(i)}(A+A')\textbf{1}}{ND^{in}_i}$ & $ANNS^{in}_i=\frac{(W+W')_{(i)}(A+A')\textbf{1}}{ND^{in}_i}$ \\
\midrule
Clustering
 & $BCC^{out}_i=\frac{(A^2A')_{ii}}{ND^{out}_i(ND^{out}_i-1)}$ & $WCC^{out}_i=\frac{(Z^2Z')_{ii}}{ND^{out}_i(ND^{out}_i-1)}$ \\
 & $BCC^{in}_i=\frac{(A'A^2)_{ii}}{ND^{in}_i(ND^{in}_i-1)}$ & $WCC^{in}_i=\frac{(Z'Z^2)_{ii}}{ND^{in}_i(ND^{in}_i-1)}$ \\
 & $BCC^{cyc}_i=\frac{(A^3)_{ii}}{ND^{in}_iND^{out}_i-NB_{i}}$ & $WCC^{cyc}_i=\frac{(Z^3)_{ii}}{ND^{in}_iND^{out}_i-NB_{i}}$ \\
 & $BCC^{mid}_i=\frac{(AA'A)_{ii}}{ND^{in}_iND^{out}_i-NB_{i}}$ & $WCC^{mid}_i=\frac{(ZZ'Z)_{ii}}{ND^{in}_iND^{out}_i-NB_{i}}$ \\
\bottomrule
\multicolumn{3}{p{10cm}}{\textit{Note}: $A_{(i)}$ is the $i$th row of $A$; $Z=[W]^{[1/3]}$ stands for the matrix obtained from $W$ after raising each entry to $1/3$; $(Z)^3_{ii}$ is the $i$th entry on the main diagonal of $Z\cdot Z\cdot Z$; and, $NB_{i}$ is the number of reciprocal partners of node $i$.} \\
\end{tabular} 
\end{small}
\end{center}
\end{table}

\section*{Acknowledgments}

Giorgio Fagiolo gratefully acknowledges support by the European Union's Horizon 2020 research and innovation program under grant agreement No. 649186 - ISIGrowth. The authors thank useful comments and suggestions from Mercedes Campi and Pablo Galaso. We also thank participants in WEHIA (Universitat Jaume I Castell\'{o} de la Plana, June 2016) and in the Workshop Din\'{a}mica Econ\'{o}mica: Teor\'{i}a y Aplicaciones (Universidad de la Rep\'{u}blica, Montevideo, November 2016).

\section*{Contributions}

MD, RM, MB, and GF conceived and designed the study. MB collected the data. MD and RM analysed the data, performed the statistical analysis, created the tables and figures, and wrote the final version of the manuscript. All authors read and approved the final manuscript.

\section*{Competing interests}

The authors declare no competing financial, professional, or personal interests that might have influenced the performance or presentation of this contribution.

\bibliography{ma_biblio}

\bibliographystyle{naturemag}

\end{document}